\documentclass[a4paper,11pt]{article}
\pdfoutput=1 

\usepackage{jheppub} 
\usepackage[table]{xcolor}
\usepackage[english]{babel}
\usepackage[utf8]{inputenc}
\usepackage{amsmath}
\usepackage{graphicx}
\usepackage{amsmath}
\usepackage{indentfirst}
\usepackage{color}
\usepackage{amssymb}
\usepackage{array} 
\usepackage{float}
\newcolumntype{M}[1]{>{\centering\arraybackslash}m{#1}}
\newcolumntype{N}{@{}m{0pt}@{}}

\def\be{\begin{equation}}
\def\ee{\end{equation}}
\def\bea{\begin{eqnarray}}
\def\eea{\end{eqnarray}}

\newcommand{\mc}{\mathcal}

\def\bel#1{\begin{equation} \label{#1}}

\def\ltap{\ \raise.3ex\hbox{$<$\kern-.75em\lower1ex\hbox{$\sim$}}\ }
\def\gtap{\ \raise.3ex\hbox{$>$\kern-.75em\lower1ex\hbox{$\sim$}}\ }
\def\gl{\ \raise.5ex\hbox{$>$}\kern-.8em\lower.5ex\hbox{$<$}\ }
\def\roughly#1{\raise.3ex\hbox{$#1$\kern-.75em\lower1ex\hbox{$\sim$}}}

\def\pref#1{(\ref{#1})}

\newcommand{\comments}[1]{}

\newcommand{\ben}{\begin{enumerate}}
\newcommand{\een}{\end{enumerate}}
\newcommand{\bi}{\begin{itemize}}
\newcommand{\ei}{\end{itemize}}
\newcommand{\ba}{\begin{align}}
\newcommand{\ea}{\end{align}}

\def\beq{\begin{equation}}
\def\eeq{\end{equation}}

\title{On the Search for Low $W_0$}

\author[a,b]{Igor Broeckel,}
\author[a,b]{Michele Cicoli,}
\author[c]{Anshuman Maharana,}
\author[c]{Kajal Singh,}
\author[d]{Kuver Sinha}
\affiliation[a]{\footnotesize Dipartimento di Fisica e Astronomia, Universitá di Bologna, via Irnerio 46, 40126 Bologna, Italy}
\affiliation[b]{\footnotesize INFN, Sezione di Bologna, viale Berti Pichat 6/2, 40127 Bologna, Italy}
\affiliation[c]{\footnotesize Harish-Chandra Research Institute, HBNI, Allahabad 211019, India}
\affiliation[d]{\footnotesize Department of Physics and Astronomy, University of Oklahoma, Norman, OK 73019, USA}

\emailAdd{igor.broeckel@bo.infn.it}
\emailAdd{michele.cicoli@unibo.it}
\emailAdd{anshumanmaharana@hri.res.in}
\emailAdd{kajalsingh@hri.res.in}
\emailAdd{kuver.sinha@ou.edu}

\abstract{The magnitude of the vacuum expectation value of the Gukov-Vafa-Witten superpotential $|W_0|$ plays a central role in the phenomenology of type IIB flux compactifications. Recent analytical constructions have shown that perturbatively flat vacua can be used to obtain very low values of $|W_0|$. We present systematic algorithms to carry out exhaustive numerical searches for such vacua. We also analyse them in the statistical context, as part of the entire ensemble of type IIB flux vacua at low $|W_0|$. Our preliminary analysis indicates that these perturbatively flat vacua are statistically sparse in the whole set of vacua at low $|W_0|$ as calculated by Denef and Douglas. Two-moduli examples are used to illustrate these more general findings in specific settings. We find that these simple cases are good examples for existence proofs but they do not feature a large statistical tuning freedom for phenomenological applications.}

\begin{document} 

\maketitle
\flushbottom

\section{Introduction}
\label{Intro}

Central to the venture of string phenomenology is to carry out explicit constructions of reliable vacua which are phenomenologically viable. In this light, particularly attractive are type IIB flux compactifications \cite{Giddings:2001yu}. Here, the complex structure moduli and axio-dilaton can be stabilised by turning on background $3$-form fluxes. There are instead various scenarios for stabilising the K\"ahler moduli (see for example \cite{Kachru:2003aw, Balasubramanian:2005zx, Berg:2005yu, Westphal:2006tn, Cicoli:2008va, Cicoli:2012fh, Cicoli:2016chb, Gallego:2017dvd, Antoniadis:2018hqy, AbdusSalam:2020ywo}). The Standard Model can be realised on intersecting D-branes, branes at singularities or their F-theory generalisations \cite{Maharana:2012tu}. Cosmic inflation can also be driven by either closed or open string moduli \cite{Cicoli:2011zz, Baumann:2014nda}. Furthermore, the large number of possibilities for choosing flux quanta leads to a multitude of solutions which can provide a way to tune the parameters of the associated $4$-dimensional effective theory \cite{Bousso:2000xa, Susskind:2003kw}.

Our understanding of the physics of type IIB flux compactifications has been growing steadily. This often involves the discovery of a novel class of solutions which have some desirable property needed for the construction of string vacua. One such very interesting class has been discovered recently \cite{Demirtas:2019sip}. These vacua are in the large complex structure limit of the underlying Calabi-Yau of the compactification. They correspond to choices of flux quanta that yield a Gukov-Vafa-Witten superpotential \cite{Gukov:1999ya} which, when computed using the perturbative part of the prepotential, is a degree-$2$ homogeneous polynomial in the complex structure moduli and the axio-dilaton. As a result, at this level, these vacua have a flat direction and the expectation value of the Gukov-Vafa-Witten superpotential vanishes along the flat direction. Therefore, these vacua have been dubbed `perturbatively flat'. The flat direction is lifted when non-perturbative corrections to the prepotential are incorporated. With this, the Gukov-Vafa-Witten superpotential acquires a value which is exponentially small (at weak string coupling). 

This discovery is particularly interesting in the context of KKLT models \cite{Kachru:2003aw}. Defining as in \cite{Demirtas:2019sip} the vacuum expectation value of the Gukov-Vafa-Witten superpotential as\footnote{Unless otherwise stated, in this article we will follow all the conventions of \cite{Demirtas:2019sip}.}
\be
W_0 \equiv {\sqrt{2 \over \pi}}  \,\langle e^{\mc{K}} \int_X G_3 \wedge \Omega \rangle\,, 
\ee
where $\mc{K}$ is the K\"ahler potential for the complex structure moduli and axio-dilaton, $G_3$ is the complexified $3$-form flux, and $\Omega$ the holomorphic $3$-form of the underlying (orientifolded) Calabi-Yau $X$, controlled KKLT vacua require exponentially small values of $|W_0|$.\footnote{For a general discussion on the magnitude $W_0$ in the context of moduli stabilisation and phenomenological implications, see \cite{Cicoli:2013swa} and references therein.} This is effectively realised in perturbatively flat vacua which feature $|W_0|\sim e^{-2\pi /(c g_s)}\ll 1$ at small string coupling $g_s\ll 1$ (with $c \in \mathbb{Q}^+$). The paper \cite{Demirtas:2019sip} presented an explicit choice of flux quanta in an orientifold of the Calabi-Yau obtained by considering a degree-$18$ hypersurface in $\mathbb{CP}_{[1,1,1,6,9]}$, which yielded $|W_0| \sim 10^{-8}$ (for earlier work on obtaining low values of $|W_0|$ see for example \cite{Giryavets:2003vd, Cole:2019enn}). Not stopping at that, an explicit example with $|W_0|$ as low as $10^{-95}$ was presented in \cite{Demirtas:2021ote, Demirtas:2021nlu}. Here, important advances were made in developing K\"ahler moduli stabilisation in this context. Furthermore, \cite{Demirtas:2020ffz,Alvarez-Garcia:2020pxd} extended the method to settings with a shrinking conifold modulus, an essential ingredient of the KKLT construction for anti-brane uplifting. The generalisation to F-theory has been considered in \cite{Honma:2021klo}. 

Perturbatively flat vacua are important also for recent LVS explicit realisations of the Standard Model with D3-branes at an orientifolded dP$_5$ singularity \cite{Cicoli:2021dhg}. In these constructions the cancellation of all D7-charges and Freed-Witten anomalies forces the presence of a hidden D7-sector with non-zero gauge fluxes which induce a T-brane background suitable for de Sitter uplifting \cite{Cicoli:2015ylx}. As can be seen from equations (5.41) and (5.46) of \cite{Cicoli:2021dhg} the T-brane contribution can give a leading order Minkowski vacuum if $|W_0|$ takes a form similar to the one typical of perturbative flat vacua since $|W_0|\sim \lambda_1 \,e^{-2 \pi \lambda_2/g_s}$ where $\lambda_1$ and $\lambda_2$ are $\mc{O}(1)$ model-dependent coefficients which depend on microscopic quantities like the Calabi-Yau Euler characteristic and intersection numbers, the number of blow-up modes, gauge flux quanta and the rank of condensing gauge groups. Phenomenologically viable vacua with a de Sitter minimum and soft terms above the TeV scale can require $|W_0|$ as small as $|W_0|\sim 10^{-13}$.\footnote{Notice that very small values of $|W_0|$ are not a necessary condition for T-brane uplifting since this depends crucially on the model-dependent values of $\lambda_1$ and $\lambda_2$. In fact, \cite{Cicoli:2017shd} found explicit LVS de Sitter models with $|W_0|\sim \mc{O}(1-10)$.} All of these are significant developments in the direction of explicit constructions of fully reliable de Sitter vacua in string theory.
 
Returning to a broader discussion, phenomenological requirements will invariably lead us to specific subclasses of flux vacua (such as perturbatively flat vacua for low $|W_0|$). As we continue to examine flux vacua in detail, we will certainly discover many other interesting subclasses. Given a subclass of flux vacua, there are two important questions that are natural to ask:

\begin{itemize}
\item How does the subclass fit within the larger ensemble of the full set of vacua? More specifically, what can we say about the set from the point of view of the statistical approach to string phenomenology \cite{Douglas:2003um, Ashok:2003gk, Denef:2004ze, Denef:2004dm, Denef:2004cf} (see \cite{Marchesano:2004yn, Dienes:2006ut, Gmeiner:2005vz, Douglas:2006xy, Acharya:2005ez, Giryavets:2004zr, Misra:2004ky, Conlon:2004ds, DeWolfe:2005gy, Hebecker:2006bn, Martinez-Pedrera:2012teo, Halverson:2018cio, Betzler:2019kon, Bena:2021wyr, Krippendorf:2021uxu} for studies in various settings in this context)?

\item How can one carry out exhaustive searches which will allow us to have a complete understanding of the vacua in this set (and their physics)?
\end{itemize}

The goal of this paper is to take the first steps and develop the methods necessary to answer the above questions in the context of perturbatively flat vacua. In the process, we hope to learn some lessons which should be applicable to the study of any subclass. Apart from the general motivation, there are  interesting reasons to address both questions in the context of perturbatively flat vacua. 

Usually, finding flux vacua requires solving a coupled set of equations involving the flux quanta and the complex structure moduli. In the case of perturbatively flat vacua, there is considerable simplification. As we will review below, to find vacua one just needs to solve a set of diophantine equations involving the flux quanta (once solutions to this set are found, the vacuum expectation values of the complex structure moduli are automatically determined by a simple analytic formula). Given this simplification, perturbatively flat vacua are the ideal set to look at to develop methods for exhaustive searches for flux vacua. 

As already mentioned, perturbatively flat vacua provide a natural way to construct KKLT models and can be useful for effective T-brane uplifting in some LVS models. Thus, developing an understanding of how they fit into the full set of flux vacua (in the statistical context) is important for obtaining the statistical predictions for observables in these models. For instance, the analysis of \cite{Broeckel:2020fdz, Broeckel:2021dpz} implies that if $|W_0|$ is exponentially small in the dilaton in most vacua in a set, then the scale of supersymmetry breaking has a logarithmic distribution. The above property is true for all perturbatively flat vacua. Therefore, gaining an understanding of what fraction of the vacua at low $|W_0|$ are perturbatively flat is central to determining the distribution of the scale of supersymmetry breaking in KKLT models. The distribution of the scale of supersymmetry breaking in the landscape is of course of much interest \cite{Susskind:2004uv, Douglas:2004qg, Dine:2004is, Arkani-Hamed:2005zuc, Kallosh:2004yh, Dine:2005yq}. 

Before closing the introduction, we would like to make some comments regarding the approach that this article takes. Work on the search for flux vacua and their properties is a two step process - development of methods and then extensive numerical scan through models. The focus of the present paper is on the 
former. While we will make use of specific models to illustrate the methods,\footnote{For this we will work with models with $2$ complex structure moduli, keeping the numerics light.} we will not be carrying out any extensive numerical scans through models. In fact, we will often stop midway with the analysis
of particular models when the necessary point regarding the methods is made. We leave detailed numerical scans of models for future work \cite{future1}.

This paper is organised as follows. In Sec. \ref{Sec2} we review the main ingredients of perturbatively flat vacua, while in Sec. \ref{Sec3} we discuss their statistical significance. Sec. \ref{Sec4} provides all the details of an algorithm to perform exhaustive searches for perturbatively flat vacua for the case with $2$ complex structure moduli. In Sec. \ref{Sec5} we outline instead a more general search algorithm which is valid in principle to obtain perturbatively flat vacua for examples with an arbitrarily large number of complex structure moduli. We present our conclusions and discuss our results in Sec. \ref{Concl}. Some technical details regarding our numerical search for cases with $2$ complex structure moduli are summarised in App. \ref{AppA}. 
 
\section{A brief review of perturbatively flat vacua}
\label{Sec2}

In this section we first recapitulate some basic material on type IIB flux compactifications and then go on to review \cite{Demirtas:2019sip}. Our discussion in the first part shall be primarily to set notation and will be quite brief. We refer the reader to \cite{Hosono:1994av, Klemm:2005tw, Curio:2000sc, Dasgupta:1999ss, Giddings:2001yu} for further details.

Type IIB flux compactifications have an internal manifold that is conformally an orientifolded Calabi-Yau $X$. To describe these in the language of special geometry, one works with a symplectic basis for $H_3(X,\mathbb{Z})$, $\{A_a, B^a \}$ for $a=1,...,h^{1,2}_-(X)$ with $A_a\cap A_b=0,$ $A_a\cap B^b=\delta^{~b}_a,$ and $B^a\cap B^b=0$, and projective coordinates on the complex structure moduli, $U^a$ (in what follows, we will take $U^0 =1$). The central object is the prepotential $\mc{F}$, which is degree-$2$ and homogeneous in the projective coordinates. The period vector is given by
\be
\Pi=\left( \begin{array}{c} \int_{B^a}\Omega\\ \int_{A_a}\Omega\end{array}\right)=\left(\begin{array}{c} \mc{F}_a\\ U^a \end{array}\right)\,.
\label{per}
\ee
The flux vectors $F$ and $H$ are obtained by integrating the $3$-form field strengths of the type IIB theory over the $A_a$ and $B_a$ cycles
\be
F=\left(\begin{array}{c}\int_{B^a}F_3\\ \int_{A_a} F_3 \end{array}\right), \qquad H=\left(\begin{array}{c}\int_{B^a} H_3\\ \int_{A_a}H_3 \end{array}\right)\,.
\label{fv}
\ee
Dirac quantisation conditions require that these are integer valued. The flux superpotential, which is classically exact, is given by
\be
W=\,\sqrt{\tfrac{2}{\pi}} \left(F-\tau H\right)^T \cdot\Sigma \cdot\Pi \,,
\label{eq:wis}
\ee
where
\be
 \Sigma = 
 \begin{pmatrix}
 0 & -1 \cr
 1 & 0
 \end{pmatrix}\,,
\ee
is the symplectic matrix. The tree-level K\"ahler potential (for the complex structure moduli and the axio-dilaton) is
\be
\mc{K}=-\ln\left(-i\Pi^{\dagger}\cdot\Sigma\cdot \Pi\right)-\ln\left(-i(\tau-\bar{\tau})\right)\,.
\label{eq:wis2}
\ee
In the large complex structure limit,\footnote{For detailed studies of flux vacua in the large complex structure limit see e.g. \cite{Dimofte:2008jg, Cicoli:2013cha, Marsh:2015zoa, Honma:2017uzn, Grimm:2019ixq, Blanco-Pillado:2020hbw, Marchesano:2021gyv}.} the prepotential is a sum of perturbative terms which are at most degree-$3$ and instanton corrections, i.e $\mathcal{F}(U)= \mathcal{F}_{\text{pert}}(U)+\mathcal{F}_{\text{inst}}(U)$ with
\be
\mc{F}_{\text{pert}}(U)=-\frac{1}{3!}\, \mc{K}_{abc} U^a U^b U^c + \frac12\, {\bf{a}}_{ab} U^a U^b + b_a U^a +\xi\,,
\label{eq:LCSprepotential}
\ee
where $\mathcal{K}_{abc}$ are the triple intersection numbers of the mirror Calabi-Yau, ${\bf{a}}_{ab}$ and $b_a$ are rational, and $\xi=-\frac{\zeta(3)\chi}{2(2\pi i)^3}$, with $\chi$ the Euler number of the Calabi-Yau. The instanton corrections are
\be
\mc{F}_{\text{inst}}(U)=\frac{1}{(2\pi i)^3}\sum_{\vec{q}} A_{\vec{q}}\,e^{2\pi i \vec{q}\cdot \vec{U}}\,,
\ee
where the sum runs over effective curves in the mirror Calabi-Yau.

Supersymmetric vacua which have $W=0$ at the perturbative level of the prepotential and also have a flat direction were termed as perturbatively flat in \cite{Demirtas:2019sip}. The basic idea of \cite{Demirtas:2019sip} is that, when the instanton corrections are incorporated, the flat direction is lifted and $W$ acquires an exponentially small vacuum expectation value. Furthermore, the paper provides an explicit algorithm to obtain perturbatively flat vacua, which was stated in the form of a Lemma.
 
The statement of the Lemma is: if there is a pair $(\vec{M},\vec{K})\in \mathbb{Z}^n\times \mathbb{Z}^n$ satisfying $N_{\rm flux} \equiv -\frac12\vec{M}\cdot \vec{K}\leq Q_{\mathrm{D3}}$ ($Q_{\mathrm{D3}}$ being the D3-charge tadpole bound), such that $N_{ab}\equiv \mc{K}_{abc} M^c$ is invertible, and $\vec{K}^T N^{-1} \vec{K}=0$, and $\vec{p}\equiv {N}^{-1}\vec{K}$ lies in the K\"ahler cone of the mirror Calabi-Yau, and such that ${\bf{a}}_{ab}\,M^b$ and $b_a \,M^a$ take on values in integers; then there exists a choice of fluxes for which a perturbatively flat vacuum exists. The perturbative F-flatness conditions are satisfied along the $1$-dimensional subspace $\vec{U}=\tau \vec{p}$, on which $W_{\text{pert}}$ vanishes. The Lemma is easily verified by taking the flux vectors to be
\be
F=(\vec{M}\cdot \vec{b},\vec{M}^T\cdot {\bf{a}},0,\vec{M}^T)\,  \phantom{abc} {\rm{and}} \phantom{abc} H=(0,\vec{K}^T,0,0) \,.
\label{eq:homogeneousfluxes}
\ee
The above choice of the flux vectors is also the most general that leads to a superpotential that is a degree-$2$ homogenous polynomial in the $\left(h^{1,2}_-+1\right)$ moduli.\footnote{Degree-$2$ homogeneous flux superpotentials and associated flat directions in toroidal compactifications were discussed in \cite{Hebecker:2017lxm}.} Note that this guarantees that the F-flatness conditions imply $W=0$, and also the existence of the flat direction.

As mentioned earlier, the flat direction is lifted by the non-perturbative terms in $\mc{F}$. Choosing the axio-dilaton to be the coordinate along the flat direction, the superpotential which is effectively generated looks like
\be
\frac{W_{\mathrm{eff}}(\tau)}{\sqrt{2/\pi}}=M^a\partial_a \mc{F}_{\mathrm{inst}}=\sum_{\vec{q}} \frac{A_{\vec{q}}\,\vec{M}\cdot \vec{q}}{(2\pi i)^2}\,e^{2\pi i \tau  \vec{p}\cdot \vec{q}}\,.
\ee
The above superpotential can lead to a controlled racetrack stabilisation if the two dominant instantons (which we will call $\vec{q}_1$ and $\vec{q}_2$) satisfy $\vec{p}\cdot \vec{q_1}\approx \vec{p}\cdot \vec{q_2}$. Furthermore, stabilisation at weak string coupling requires that there is a hierarchy between the prefactors of the instantons. This amounts to a hierarchy in the associated Gopakumar-Vafa invariants \cite{Gopakumar:1998ii, Gopakumar:1998jq}.

\section{Expectations from statistics}
\label{Sec3}

As discussed in the introduction, it is of much interest to develop an understanding of how perturbatively flat vacua fit in the larger ensemble of type IIB flux vacua in the statistical sense. The question is central to understanding the distribution of the scale of supersymmetry breaking for KKLT vacua \cite{Broeckel:2020fdz}. Perturbatively flat vacua are supersymmetric (even after the incorporation of instanton effects in the prepotential) and have low values of $|W_0|$. The statistical properties of such vacua were derived in \cite{Denef:2004ze}. The number of such vacua $\mc{N}$ with the value of $|W_0|^2$ below $\lambda_*$ is given by an integral of a density over the moduli space\footnote{In the discussion below, we translate the results of \cite{Denef:2004ze} and report them in the conventions of \cite{Demirtas:2019sip}.}
\be
\mc{N}(N_{\rm flux} \leq Q_{\rm D3},|W_0|^2 \leq \lambda_*)=\frac{(2\pi Q_{\rm D3} )^{2m} \pi \lambda_*}{2 (2m)!}  \int_\mc{M} d^{2m} z\hspace{2pt} \sqrt{g} \hspace{2pt} \rho(z)\,,
\label{N}
\ee
where the density function is given by
\be
\rho(z)=\frac{2\pi m}{\pi^{2m}Q_{\rm D3}}\,I(\mc{F}) \phantom{abc}  {\rm{for}} \phantom{abc} 
I(\mc{F})=\int d^{2 h^{1,2}_-} Z\,e^{-|Z|^2} |\rm{det} 
\left(\begin{matrix} 
0 & Z_J \\
Z_I & e^{\mc{K}} \mc{F}_{IJK} \bar{Z}^K
\end{matrix}
\right)|^2,
\label{rz}
\ee
with $m = h^{1,2}_-+1$ ($h^{1,2}_-$ being the number of complex structure moduli). The $d^{2m}z$ integration runs over the $2m$-dimensional space of the complex structure moduli and the axio-dilaton and it involves its metric. $\mc{F}_{IJK}$ are components of triple derivatives of the prepotential expressed in a local frame. The integration variables $Z_I$ are related to derivatives of the flux superpotential, but can be thought of as dummy integration variables for the purposes of computation of $I(\mc{F})$. 
 
Now, let us turn to examining perturbatively flat vacua in this context. For this, we will exploit universal properties of these vacua. A striking property of these vacua is that for all of them, at their minima
\be
\vec{U} = \tau \vec{p}\,,
\label{Ulocus}
\ee
where the vector $\vec{p}$ is real and has all positive entries. The real parts of $\vec{U}$ are axionic. Thus, after the axions are brought to their fundamental domain, the relations in \pref{Ulocus} will continue to hold modulo factors of integers. Therefore, the solutions under consideration are contained in a subspace of the moduli space which is isomorphic to $\mc{M}_\tau \times \left( {\mathbb{R}}^+ \right)^{h^{1,2}_-}$. Given that perturbatively flat vacua are a subset of the set of all solutions satisfying (\ref{Ulocus}), a necessary criterion for them to have statistical significance is that the set of all solutions on the subspace $\pref{Ulocus}$ have statistical significance. This question can be examined from the point of view of the densities of \cite{Denef:2004ze}. Notice that the entire moduli space is $2\left(h^{1,2}_-+1\right)$-dimensional, while the subspace described by (\ref{Ulocus}) is only $\left(h^{1,2}_-+2\right)$-dimensional. Since the subspace (\ref{Ulocus}) is of lower dimensionality than the entire moduli space, and the densities are smooth function on the entire moduli space, even the set of all vacua on this subspace are not expected to be of statistical significance, implying the same for perturbatively flat vacua. Hence perturbatively flat vacua are expected to be statistically sparse in the set of flux vacua with low $|W_0|$ as given by the distribution from \cite{Denef:2004ze} which already gives a much smaller number of vacua with respect to cases with $|W_0|\sim \mc{O}(1-10)$ since (\ref{N}) is linear in $\lambda_*$, implying (at fixed $Q_{\rm D3}$) $\mathcal{N}(|W_0|^2 \leq \lambda_*) / \mathcal{N}(|W_0|^2 \leq 1) \sim \lambda_*$. 

Notice that comparisons in \cite{Denef:2004ze} of results of explicit searches to the predictions making use of densities did show local fluctuations (such as overdensities and voids) but these were local effects not having any effect on the overall statistical predictions. It is however important to check if the distributions of \cite{Denef:2004ze} are peaked along the space described by \pref{Ulocus}. Consider the subspace in which the axio-dilaton is purely imaginary. Being on \pref{Ulocus}, then implies that $U^a$ are also purely imaginary. The densities of \cite{Denef:2004ze} can be expressed in terms of the K\"ahler potential, the metric on the moduli space and triple derivatives of the prepotential $\mc{F}_{IJK}$. For the perturbative part of the prepotential (which dominates in the large complex structure limit), all the above quantities are independent of the real part of $U^a$. Thus the densities are also independent of the real part of $U^a$. Therefore, at fixed purely imaginary axio-dilaton, moving away from the locus $\pref{Ulocus}$ by switching on a non-zero real part of $U^a$ does not lead to a fall in the value of the density. Similar considerations also apply when the axio-dilaton is not purely imaginary. 

We would like to close this section with a cautionary remark. The diagnostics presented here relies on the fact that the basic assumption of \cite{Denef:2004ze} is valid, i.e. that the space of flux vacua of a given compactification can be described by smooth density functions obtained by replacing sums over flux quanta by integrals. If for some reason this fails, the diagnostics would be irrelevant. Next, we turn our discussion of setting up exhaustive searches for perturbatively flat vacua, which is crucial for developing a full understanding of their properties.

\section{Exhaustive search with two moduli}
\label{Sec4}

\subsection{The $\mathbb{CP}_{[1,1,1,6,9]}$ example}
\label{cp}

In this section we describe algorithms for carrying out exhaustive searches for perturbatively flat vacua in Calabi-Yau threefolds with $2$ complex structure moduli. As mentioned in the introduction, even if this paper intends mainly to focus on methods for searches of flux vacua, for completeness we will present  an explicit example in full detail. We do so by looking at the degree-$18$ hypersurface in $\mathbb{CP}_{[1,1,1,6,9]}$ used in \cite{Demirtas:2019sip} (studied in the context of mirror symmetry in \cite{Candelas:1994hw}).

We begin by recalling some basic facts about the Calabi-Yau and some details of the analysis of \cite{Demirtas:2019sip}. The Calabi-Yau has $272$ complex structure moduli but has a $\mc{G}=\mathbb{Z}_6\times \mathbb{Z}_{18}$ symmetry. By considering $\mc{G}$-invariant fluxes, one is guaranteed to stabilise on the $\mc{G}$-symmetric locus (see \cite{Giryavets:2003vd}). Thus the stabilisation problem is effectively reduced to a $2$-moduli one. The relevant geometric data are
\be
\mc{K}_{111}= 9\,,\quad \mc{K}_{112}=3\,,\quad \mc{K}_{122}=1\,,\quad {\bf{a}}= \frac12 
\begin{pmatrix}
9 & 3 \\
3 & 0
\end{pmatrix}\,,\quad \vec{b}=\frac14
\begin{pmatrix}
17\\
6
\end{pmatrix}\,,
\label{eq:geodata}
\ee
and the instanton corrections are
\begin{align}
\label{eq:theinst}
& (2\pi i)^3 \mc{F}_{\text{inst}}=\mc{F}_1 + \mc{F}_2 + \cdots\, , \\
& \mc{F}_1 =  - 540\, q_1 - 3\,q_2\, ,\label{eqf1} \\
& \mc{F}_2 = -\frac{1215}{2}\, q_1^2 +1080\,q_1 q_2 + \frac{45}{8}\,q_2^2\, ,\label{eqf2}
\end{align}
where $q_a=\exp(2\pi i U^a)$ with $a \in\{1,2\}$. We consider the orientifold involution described in \cite{Louis:2012nb} which yields a D3-charge $Q_{\rm D3}=138$. 

Making use of \eqref{eq:geodata}, the condition $\vec{K}^T {N}^{-1} \vec{K} = 0$ gives
\be
M^1 = \frac{M^2 K^2 \left(2K^1 - 3K^2\right)}{\left(K^1 - 3K^2\right)^2}\,,
\label{eq:lemmaeq}
\ee
and the flat direction is
\be
\label{usol}
\vec{U} = \tau \begin{pmatrix}
p^1\\
p^2
\end{pmatrix} = \frac{\tau \left(K^1 - 3K^2\right)}{M^2} \begin{pmatrix}
- K^2/K^1\\
1
\end{pmatrix}\,.
\ee
The following choice of the vectors $(\vec{M}, \vec{K})$ 
\be
\label{fsol}
\vec{M}=\begin{pmatrix}
-16\\
\phantom{-}50
\end{pmatrix}\, ,\quad \vec{K}=\begin{pmatrix}
\phantom{-}3\\
-4
\end{pmatrix}\, ,
\ee
meets all the conditions of the Lemma and the flat direction can be lifted by the inclusion of non-perturbative terms.

In the large complex structure limit, the K\"ahler potential (for the complex structure moduli and axio-dilaton) is given by 
\be
K = - \ln \left( i {1 \over 6} \mc{K}_{abc} (U^a - \bar{U}^a) (U^b - \bar{U}^b) (U^c - \bar{U}^c) + 4 i \xi \right) - \ln \left( -i(\tau - \bar{\tau}) \right).
\ee
We are interested in the locus $U^a = p^a \tau$. Furthermore, since in this limit ${\rm{Im}} (U^a) >1$, the term involving $\xi$ is subdominant. Thus, along this locus one has
\begin{eqnarray}
\label{kahpot}
K &=& - \ln \left({1 \over 6} \mc{K}_{abc} p^a p^b p^c \left( -i(\tau - \bar{\tau})\right)^{3} + 4 i \xi \right) - \ln \left( -i(\tau - \bar{\tau}) \right) \cr
  &\sim&  - \ln \left(  {1 \over 6} \mc{K}_{abc} p^a p^b p^c \right) - 4 \ln \left( -i(\tau - \bar{\tau}) \right). 
\end{eqnarray}
The effective superpotential for stabilising the perturbatively flat direction takes the form
\be
W_{\text{eff}}(\tau)=c\,\left(e^{2\pi i p^1 \tau}+A e^{2\pi i p^2 \tau}\right)\, + \cdots \,,
\ee
where $c=\sqrt{\frac{2}{\pi}}\frac{8640}{(2\pi i)^2}$ and $A=-\frac{5}{288}$. Making use of the fluxes in \pref{fsol}, it can be easily found that $|W_0| \simeq 2 \times 10^{-8}$.

\subsection{The algorithm}
\label{algo}

Now, we describe an algorithm for finding all perturbatively flat vacua in the $\mathbb{CP}_{[1,1,1,6,9]}$ model, which can however be easily generalised to other $2$-moduli examples. The F-flatness condition is $D_\tau W_{\rm{eff}} = \left( \partial_\tau + \partial_\tau K \right) W_{\rm eff} = 0$. Note that the form of the K\"ahler potential \pref{kahpot} implies that $\partial_\tau K \propto  \left({\rm Im} \tau\right)^{-1} = g_s$. Therefore, for consistent stabilisation at weak string coupling, the term involving $\partial_\tau K$ must be a small correction to the F-flatness condition. The F-flatness condition neglecting this term is
\be
e^{ 2 \pi i \tau (p^1 - p^2)}  = - {A p^2 \over p^1} \,.
\label{fsol}
\ee
Let us start our search by considering the cases with $p^1 > p^2$. Now, by making use of \pref{usol} the condition $p^1 > p^2$ translates to
\be
- {K^2 \over K^1} > 1 \,.
\ee
Thus $K^1$ and $K^2$ have to be of opposite sign. Furthermore, the entire set of conditions in the Lemma have a symmetry:
\be
\vec{K} \to - \vec{K} \phantom{abc} \textrm{and} \phantom{abc} \vec{M} \to - \vec{M}\,.
\ee
This in fact corresponds to an S-duality transformation with the centre of the group. Thus, without loss of generality, we will look at cases with $K^1 > 0$ and $K^2 <0$. With this, the factor $\left(K^1 - 3K^2\right)$ in \pref{usol} is positive, implying that $M^2$ must be positive (so that $p^2$ is positive, as required by the K\"ahler cone condition in the Lemma). With these signs of $K^1$, $K^2$ and $M^2$, equation \pref{eq:lemmaeq} gives the sign of $M^1$ to be 
negative. This implies that $A=M^2/(180 M^1)$ has to be negative, which can be compatible with (\ref{fsol}) if at the minimum ${\rm Re}(\tau)= k/(p^1-p^2) \phantom{a} {\rm{mod}} \phantom{a} \mathbb{Z}$ with $k\in\mathbb{Z}$. The above suggests the following efficient algorithm to carry out an exhaustive search for vacua:

\begin{enumerate}
\item Consider a rational number $x$ between $0$ and $1$, express this as $x = {p / q}$ such that $p$ and $q$ are positive and have no common factors. Define the vector
\be
\vec{\tilde{K}}=\begin{pmatrix}
\tilde{K}^1 \\
\tilde{K}^2
\end{pmatrix}=\begin{pmatrix}
p \\
-q
\end{pmatrix}\,.
\ee
The vector $\vec{\tilde{K}}$ will eventually be related to the vector $\vec{K}$ being searched for.

\item Now, compute the ratio
\be
y = \frac{ \tilde{K}^2 \left(2\tilde{K}^1 - 3\tilde{K}^2\right)}{\left(\tilde{K}^1 - 3\tilde{K}^2\right)^2}\,.
\label{ydef}
\ee
Note that this is related to the ratio $M^1 / M^2$ as given by \pref{eq:lemmaeq}. Express $y$ as $r/s$, such that $r$ and $s$ have no common factors, and $s>0$. Define
\be
\vec{\tilde{M}}  =\begin{pmatrix}
r \\
s
\end{pmatrix}\,.
\ee
The vector $\vec{\tilde M}$ will be eventually related to the vector $\vec{M}$ being searched for.

\item Check if $K_{abc} \tilde{M}^c$ is invertible or not. If it is not invertible, discard $x$ and start again with a new one. If it is invertible, then proceed further.

\item Compute the values 
\be
\alpha {\bf{a}} \,.\, \vec{\tilde{M}}   \phantom{abc} \textrm{and} \phantom{abc}  \alpha \vec{b}\, .\, \vec{\tilde{M}}\,,
\ee
for $\alpha = 1,2,4$. Determine the minimum value of $\alpha$ for which the above quantities are integer valued. Call this $\hat\alpha$. Note that they certainly must be integer valued for the case of $\alpha =4$, given the form of $\bf{a}$ and $\vec{b}$ in equation \pref{eq:geodata}.

\item Consider the quantity
\be
- {1 \over 2} \hat\alpha  \vec{\tilde{M}} \, . \,\vec{ \tilde{K}}\,.
\ee
If this does not satisfy the D3-tadpole bound, then discard $x$ and move to another $x$. If it lies in the allowed range, we have a solution satisfying all conditions of the Lemma with 
\be
\vec{M} = \hat\alpha  \vec{\tilde{M}}   \phantom{abc} \textrm{and} \phantom{abc} \vec{K} = \vec{\tilde{K}} \,.
\ee
Also, for {\it{any}} positive integer $\beta$ such that $- {1 \over 2} \beta  \hat\alpha  \vec{\tilde{M}} .\tilde{K}$ satisfies the D3-tadpole bound, we have solutions
\be
\vec{M} = \hat\alpha \beta_1 \vec{\tilde{M}}   \phantom{abc} \textrm{and} \phantom{abc} \vec{K} = \beta_2 \vec{\tilde{K}}\,,
\ee
where $\beta_1$ and $\beta_2$ are positive and provide a factorisation of $\beta$.

\item To scan through all $x$, note that the signs of $K^a$ and $M^a$ (with our working assumption of $K^1 >0$) are such that $M^1 K^1 < 0$ and $M^2 K^2 < 0$. Thus both terms contribute with a {\it positive} sign to the inner product $-{1 \over 2} \vec{M}. \vec{K}$. Therefore, {\it the maximum value} of $|K^2|$ necessary to carry out an exhaustive search is $2Q_{\rm D3} =  2 \times 138$ (as higher values would violate the D3-tadpole condition). This bound on $|K^2|$ implies that we need to consider only those $x$ for which $q \leq 2  Q_{\rm D3}$. Reduced rationals between $0$ and $1$ with a fixed upper bound on the denominator are given by the Farey sequence. Thus an exhaustive search is carried out by selecting $x$ from the set $\rm{Farey}_{2Q_{\rm D3}}$.

\item Scan through the solutions obtained in this way, checking that non-perturbative effects lead to stabilisation at weak string coupling. Discard the ones that do not satisfy this condition.

\item Enlarge the solution list by considering the solutions obtained by the above process and then generating the solutions related to them by the S-duality symmetry
\be
  \vec{K} \to - \vec{K}    \phantom{abc} \textrm{and} \phantom{abc} \vec{M} \to - \vec{M}\,.
\ee

\item Finally, run the same algorithm considering the possibility of $p^1 \leq p^2$.
\end{enumerate}

Carrying out the search using the above algorithm, after S-duality identification, we find that there exist only $2$ solutions which satisfy the conditions of the Lemma, although one of them does not have a very low value of $|W_0|$ since it features just $|W_0|\simeq 0.3$. We report these in Tab. \ref{Tab1} along with the associated value of $|W_0|$, after stabilisation by non-perturbative effects (the second entry in the table is the solution reported in \cite{Demirtas:2019sip}). We conclude that the $\mathbb{CP}_{[1,1,1,6,9]}$ model essentially features only $1$ perturbatively flat solution with very low $|W_0|$. 

\begin{table}
\centering
\begin{tabular}{ | p{1.5cm} | p{1.5cm}  | p{1.0cm}  | p{1.5cm} |p{1.0cm} | p{2.0cm} | p{2.5cm} | }
\hline
$ \vec{M}^T$ & $\vec{K}^T$& $\vec{b}.\vec{M}$ & $({\bf{a}}.\vec{M})^T$& $N_{\rm flux}$& $\tau$ & $|W_0|$ \\
\hline
(32, -98) & (-1, 2) &-11 & (-3, 48) & 114 & 4.884 $i$ & $0.2871$\\
\hline
(16, -50) & (-3, 4)  & -7 & (-3, 24)&124 & 6.855 $i$ & $2.048\times 10^{-8}$ \\
\hline
\end{tabular}
\caption{All perturbatively flat vacua for the $\mathbb{CP}_{[1,1,1,6,9]}$ example.}
\label{Tab1}
\end{table}

\subsection{General treatment of $2$-moduli case}
\label{gtm}

In this section we will present a general discussion of the cases with $2$ complex structure moduli. A key-feature of the algorithm in Sec. \ref{algo} was the bound on the range of the elements of the vectors $\vec{M}$ and $\vec{K}$. First we show that this follows from general considerations. The definition $\vec{p} = N^{-1} \vec{K}$ , together with the equation $\vec{K}^T N^{-1} \vec{K} = 0$ implies
\be
\vec{K}^T \vec{p} =0\,.
\ee
The requirement that $\vec{p}$ lies in the K\"ahler cone, then implies that $K^1$ and $K^2$ have opposite signs. By making use of the definition of $\vec{p}$ again, the equation $\vec{K}^T N^{-1} \vec{K}$ can alternatively be written as  
\be
p^a p^b K_{abc} M^c = 0\,.
\ee
The requirement that $\vec{p}$ lies in the K\"ahler cone implies also that $\tilde{p}_c \equiv p^{a} p^{b} K_{abc}$ has positive entries. Thus the vector $\vec{M}$ satisfies an equation similar to $\vec{K}$, i.e.
\be
\vec{M}^T \vec{\tilde{p}} = 0\,.
\ee
Therefore, $M^1$ and $M^2$ have to have different signs. 

Now, if $K^1$ and $M^1$ have the same sign, then so would $K^2$ and $M^2$. And this would imply a negative value for $N_{\rm flux} = - {1 \over 2} \vec{M} . \vec{K}$, which is impossible for imaginary self dual fluxes.\footnote{Any fluxes that solve the conditions being imposed are imaginary self dual from the $10$-dimensional perspective (see e.g. \cite{Giddings:2001yu}).} Thus viable solutions feature $K^1$ and $M^1$ of opposite sign. This implies that both terms contributing to the $N_{\rm flux}$ inner product have to be positive. Thus, an exhaustive search can be carried out by considering the range
\be
|M^a| \leq 2 Q_{\rm D3} \phantom{abc}  {\rm{and}} \phantom{abc} |K^a| \leq 2 Q_{\rm D3}\,,
\label{genbou}
\ee
which is the same as for the $\mathbb{CP}_{[1,1,1,6,9]}$ example, obtained by using slightly different considerations. As an example, we have carried out the analysis for the Calabi-Yau embedded in $\mathbb{CP}_{[1,1,2,2,2]}$ discussed in \cite{Candelas:1993dm}. To gain a model-independent picture, we treat $Q_{\rm D3}$ as a free parameter. The results are summarised in Tab. \ref{Tab2}. 

\begin{table}
\centering
\begin{tabular}{ | p{3.5cm}  | p{6.5cm}  | }
\hline
$\quad \quad \quad \quad Q_{\rm D3}$ & $\,\,$Number of perturbatively flat vacua \\
\hline
$\quad \quad \quad \quad \,50$ & $\quad \quad \quad \quad \qquad \quad \,37$ \\
\hline
$\quad \quad \quad \quad 100$ & $\quad \quad \quad \quad \qquad \quad 128$\\
\hline
$\quad \quad \quad \quad 250$ & $\quad \quad \quad \quad \qquad \quad 531$\\
\hline
$\quad \quad \quad \quad 500$ & $\quad \quad \quad \quad \qquad \quad 1445$\\
    \hline
\end{tabular}
\caption{Number of perturbatively flat vacua in the $\mathbb{CP}_{[1,1,2,2,2]}$ model taking $Q_{\rm D3}$ as a free parameter. The reported numbers are before imposing any of the following $3$ requirements: stabilisation at weak string coupling, low $|W_0|$, S-duality identification.}
\label{Tab2}
\end{table}

All these solutions can potentially correspond to perturbatively flat vacua but these numbers would be reduced by the following $3$ requirements which have still to be imposed: ($i$) dilaton stabilisation at weak string coupling by instanton effects; ($ii$) a value of $|W_0|$ which is indeed very small (i.e. not of order $|W_0|\simeq 0.3$ as for $1$ solution in the $\mathbb{CP}_{[1,1,1,6,9]}$ example); ($iii$) possible equivalences between solutions via S-duality. Given that these numbers are still small to be attractive in the context of a landscape, we do not push the analysis further.
  
It is important to note that the key-element in obtaining the bounds in \pref{genbou} was the sign correlations between the elements in $\vec{M}$ and $\vec{K}$. While the arguments in the first part of this section hold for any number of moduli, it is easy to see that the sign correlations need not hold when there are more than $2$ moduli. To remedy this, we will discuss a more general method in Sec. \ref{Sec5}.

\subsection{Comparison with statistics}
 
Let us compare our results with the statistical expectations of \cite{Denef:2004ze}. For $h^{1,2}_-= 2$, \pref{N} yields
\be
\label{statfin}
\mc{N}(N_{\rm flux} \leq Q_{\rm D3},|W_0|^2\leq \lambda_*)=\left(\frac{2^6\pi^4}{5!}\right) Q_{\rm D3}^5 \lambda_*
\int_\mc{M} d^6z \sqrt{g} \hspace{3pt} e^{2\mc{K}} \mc{F}_{abc}\bar{\mc{F}}^{abc} \,,
\ee 
where the indices of $\mc{F}$ have been converted to tangent bundle ones. For the $\mathbb{CP}_{[1,1,1,6,9]}$ example discussed in Sec. \ref{cp}, carrying out the integration over the large complex structure patch one finds
\be
\mc{N} (N_{\rm flux} \leq Q_{\rm D3}=138,|W_0|^2\leq \lambda_*)\simeq 3 \times 10^{12} \lambda_*\,.
\ee
As pointed out in \cite{Demirtas:2019sip}, this predicts  the lowest value of $|W_0|$ being of order $6 \times10^{-7}$, close to what was found. On the other hand, the same formula predicts $\mc{O}(10^8)$ vacua for $|W_0| \lesssim 0.01$, even if our exhaustive search has shown that there is only $1$ vacuum with such a feature, in agreement with the argument presented in Sec. \ref{Sec3}.\footnote{In this context, we would like to mention that the values of $|W_0|$ obtained after stabilisation crucially depend on the hierarchy in the Gopakumar-Vafa invariants. However, the densities of \cite{Denef:2004ze} in the moduli space in the large complex structure limit have mild sensitivity to this. This is in keeping with the arguments of Sec. \ref{Sec3} which suggest that perturbatively flat vacua are a small fraction of the vacua at low $|W_0|$.} We therefore conclude that in the $\mathbb{CP}_{[1,1,1,6,9]}$ model, perturbatively flat vacua are interesting examples to show explicitly the existence of vacua with very low $|W_0|$, but they do not possess any tuning freedom in the value of $|W_0|$. Given the argument presented in Sec. \ref{gtm}, we expect this conclusion to hold for all cases with $2$ complex structure moduli. Notice, for example, that in the $\mathbb{CP}_{[1,1,2,2,2]}$ model the number of perturbatively flat vacua summarised in Tab. \ref{Tab2} is also much less than as predicted by the $Q_{\rm D3}^5$ scaling of \pref{statfin}. Models with more than $2$ complex structure moduli require a refined analysis for exhaustive searches which we outline in the next section, although the analysis of Sec. \ref{Sec3} indicates that they should still be statistically sparse.

Let us close this section by stressing that a key-assumption in the derivation of the results of \cite{Denef:2004ze}, is a high density of flux vacua allowing for the sums over integer fluxes to be converted to integrals. Our results indicate that for the $\mathbb{CP}_{[1,1,1,6,9]}$ model, under these circumstances, there are many more vacua at low $|W_0|$ that remain to be discovered.

\section{A more general search algorithm}
\label{Sec5}

The key to carry out exhaustive searches is isolating the region in the flux vector space which contains all perturbatively flat vacua. Once such a region is obtained, one can carry out numerical searches in this region to obtain all solutions (if the region is not too large). In this section we present a general method to isolate such regions which is in principle valid for examples with an arbitrary large number of complex structure moduli. Here, we will discuss the method and leave its detailed numerical implementation for future work.\footnote{Our preliminary analysis indicates that the numerics can be quite involved when one considers models with more than $2$ moduli.}

Central to our arguments will be certain properties of $N_{\rm flux}$. Recall that the quantity $- {1 \over 2} \vec{M} . \vec{K}$ is equal to the contribution of the fluxes to the D3-charge
\be
N_{\rm flux} =  -\frac12\, \vec{M} . \vec{K} = {1 \over { (2 \pi)^4 \alpha'^2 } } \int_X H_3 \wedge F_3\,,
\label{nd3}
\ee
where the integration is over the Calabi-Yau $X$. The fluxes of interest to us correspond to an imaginary self dual $G_3$, i.e.
\be
 * \frac{H_3}{g_s} =  - \left( F_3 - C_0 H_3 \right) .
\label{isd}
\ee
Thus (see e.g. \cite{Kachru:2002he})
\be
\int_X H_3 \wedge F_3= \frac{1}{3! g_s} \int_X d^6 y \sqrt{g_6}\, H_3^2  \,.
\label{flpos}
\ee
This is the usual argument given to show that $N_{\rm flux}$ is positive semi-definite. Here we list two consequences that are important for our arguments:
\begin{enumerate}
\item[(a)] Equation (\ref{flpos}) implies that the only way for $N_{\rm flux}$ to vanish is $H_3 = 0$. Equation \pref{isd} then implies that $F_3 = 0$. Translating this in terms of the vectors $\vec{M}$ and $\vec{K}$, one learns that, for consistent solutions, $N_{\rm flux} = 0$ only if $\vec{M} = \vec{K} = 0$.

\item[(b)] The derivation of \pref{flpos} does not make use of flux integrality. Thus, the conclusions of the above point remain valid even when one considers fluxes which do not obey the Dirac quantisation conditions (we will do so as an intermediate step in our analysis).
\end{enumerate}

Now, returning to finding the solutions to the conditions of the Lemma, let us think of carrying out a search by scanning through the vectors $\vec{M}$ and $\vec{K}$, by starting from the origin and progressively going through points with larger and larger $|\vec{M}|$ and $|\vec{K}|$. We would like to obtain upper bounds on the values of $|\vec{M}|$ and $|\vec{K}|$ which can possibly yield solutions to the conditions of the Lemma. For this, we write the D3-tadpole condition as
\be
- {1 \over 2} | \vec{M} | | \vec{K} | \epsilon \leq Q_{\rm D3}\,,
\label{dtad}
\ee
where $\epsilon$ is the cosine of the angle between the vectors $\vec{M}$ and $\vec{K}$. Since both $|\vec{M}|$ and $|\vec{K}|$ are bounded from below, the only way $|\vec{M}|$ or $|\vec{K}|$ (or both) can be large is if $|\epsilon|$ is small. While in general the cosine of the angle between two vectors in $\mathbb{Z}^n$ can be arbitrarily small, our interest is only in vectors that satisfy the conditions of the Lemma (i.e. provide consistent solutions to the type IIB equations of motion). We begin by defining
\be
\hat{m} = \frac{\vec{M}}{|\vec{M}|}\,,  \phantom{abc} \hat{k} = \frac{\vec{K}}{|\vec{K}|} \phantom{abc}  {\rm{and}} \phantom{abc} 
{\bf{\hat{n}}}_{ab} = K_{abc} m^c\,.
\label{newdef}
\ee
The vectors $\vec{m}$ and $\vec{k}$ lie on the unit sphere and the integrality condition of the fluxes is now that the ratio of any two components of the vectors is rational. The equation constraining the vectors in the Lemma becomes
\be
\hat{k}^T \,{\bf{\hat{n}}}\, \hat{k} = 0\,.
\label{lemeqq}
\ee
We will consider the equation \pref{lemeqq} as an equation over {\it {real}} variables $\vec{m}$ and $\vec{k}$ (taking values on the unit sphere). Furthermore, we will demand that the vector $\hat{p} = {\bf{\hat{n}}}^{-1}\, \hat{k}$ lies in the K\"ahler cone of the mirror Calabi-Yau. With the variables taking on values over reals, the solution space can be studied using standard numerical methods. A lower bound on $|\epsilon|$ can be obtained by numerically searching for the minimum (or infimum) of $|\hat{m}. \hat{k}|$ in the solution space. Once such a bound is obtained, an exhaustive search can be carried out
by scanning through
\be
0 < |\vec{M}|, |\vec{K}| \leq { 2Q_{\rm D3}   \over |\epsilon|_{\rm inf} }\,.
\label{range}
\ee
We note that a bound so obtained is conservative, due to the expansion of the domain of the variables to the reals.

Next, we would like to discuss some aspects of the minimisation problem at hand. As we have reviewed above, as long as one is in the physically allowed region of the moduli space, $N_{\rm flux}$ is always greater than zero, i.e. $|\epsilon| > 0$. Thus there are two possibilities for the infimum of $|\epsilon|$: either it is a positive number or it is equal to zero. In the former case, an exhaustive search can be carried by considering $|\vec{M}|$ and $|\vec{K}|$ in the range (\ref{range}).

The later case (in which $|\epsilon|$ takes on arbitrarily small values) is more subtle. In this case, there would be a point with $|\epsilon| = 0$ as a limit point of points in the solution space. Since all points in the physically allowed region must have $|\epsilon| > 0$, the limit point must lie in the boundary of the physical region. Typically, as one approaches the boundary, one loses control over the effective field theory or encounters phenomenological challenges. Taking this into consideration will lead to an effective $|\epsilon|_{\rm inf}$ which can be used to determine a region to carry out exhaustive searches.\footnote{In principle, there can be situations where there are no good reasons to exclude a region of arbitrary small $\epsilon$. In such a case, one would need to carry out a more extensive search along vectors $\vec{m}$ and $\vec{k}$ in this region.}

To illustrate the method in a concrete setting, we consider the $39$ Calabi-Yau threefolds with $2$ K\"ahler moduli\footnote{The mirrors have $2$ complex structure moduli. Of course, our analysis in Sec. \ref{Sec4} already provides regions for exhaustive searches for these. The goal here is to obtain the analogous regions from the algorithm presented in this section. We will proceed without worrying about issues that can arise from orientifolding.} constructed by Kreuzer and Skarke in \cite{Kreuzer:2000xy} and listed (along with the intersection numbers) in Table $11$ of \cite{AbdusSalam:2020ywo}. In all these cases we have followed the above described procedure to determine $|\epsilon|_{\rm inf}$. For $22$ of them, $|\epsilon|_{\rm inf}$ does not take values close to zero, implying a strong bound on the region where all solutions are contained. We record the associated values of $|\epsilon|_{\rm inf}$ for them in Tab. $3$ in App. \ref{AppA}.

On the other hand, in the remaining $17$, the numerics yield very low values of $|\epsilon|_{\rm inf}$. Thus these models might seem to be more promising to find a larger number of perturbatively flat vacua (from the perspective of the present algorithm). However, as we discuss below, most of the would-be solutions would not be ideal for phenomenological applications. In fact, in these cases we find a solution to the equations with $|\epsilon| = 0$ on the boundary of the K\"ahler cone, i.e. $\hat{p}^b = 0$ for some $b$. The definition of $\bf{\hat{n}}$ in \pref{newdef} together with the definition of $\hat{p}$ implies that $\vec{p} = \hat{p}\, { |\vec{K}| \over | \vec{M}|}$. Thus we have the relation
\be
 \vec{U}  = \hat{p} \,{ |\vec{K}| \over | \vec{M}|}\,\tau\,.
\ee
Being in the large complex structure limit requires ${\rm Im} ( U^c) > 1$ $\forall c$. In the limit where $\hat{p}^b \to 0$ for one of the $b$,\footnote{In none of the $17$ cases both $\hat{p}^1$ and $\hat{p}^2$ tend to zero.} this can be achieved by either $ { |\vec{K}| \big{/ }| \vec{M}|}\gg 1$ or ${\rm Im}(\tau)\gg 1$. However both cases are problematic for the following reasons. ${ |\vec{K}| \big{/ }| \vec{M}|}\gg 1$  will induce large hierarchies in the vector $\vec{p}$, making it unsuited for racetrack stabilisation at small string coupling. On the other hand, ${\rm Im}(\tau)=g_s^{-1}$ cannot become too small without inducing phenomenological problems. In fact, in type IIB compactifications the Standard Model can either be realised on D3- or D7-branes. In the first case, the strength of the gauge couplings is set by $g_s$, and in the second by the Einstein frame volume of the $4$-cycle wrapped by the D7-stack (which we denote as ${\rm Re}(T_{\rm SM})$). In the scenario at hand, however K\"ahler moduli stabilisation \cite{Demirtas:2021nlu} gives
\be
\frac{4\pi}{g_{\rm SM}^2} = {\rm Re}(T_{\rm SM}) \simeq \frac{1}{2\pi} \ln |W_0|^{-1} \sim \frac{1}{g_s}\,.
\label{couplings}
\ee
Thus, irrespective of how the Standard Model is realised, in perturbatively flat vacua the strength of its gauge couplings is always determined by $g_s$.\footnote{Notice that in the string frame $\left.{\rm Re}(T_{\rm SM})\right|_{\rm str} = g_s {\rm Re}(T_{\rm SM})\sim \mc{O}(1)$, implying that one should consider all perturbative and non-perturbative $\alpha'$ corrections at string tree-level, except when $|W_0|\ll 1$ for those which come from $10$-dimensional terms proportional to $G_3^{2n}$ with $n>1$ \cite{Cicoli:2021rub}. However, as shown in \cite{Demirtas:2021nlu}, these $\alpha'$ effects should induce just a subdominant shift of the KKLT minimum for $|W_0|\ll 1$.} This effectively sets a lower bound on the range of interest for $g_s$ (for instance one can demand $10^{-3} \lesssim g_s \lesssim 0.1$). Thus, the regions of small $\epsilon$ should be effectively avoided, implying that also the remaining $17$ models are not expected to produce a large number of perturbatively flat vacua which are phenomenologically viable.

Before closing this section, we note that the key-aspect of the algorithm has been that, by determining the minimum value of the angle between the flux vectors, one can isolate a region by scanning through which exhaustive searches can be carried out. It will be interesting to see if the same considerations
can be used in other settings.

\section{Conclusion and discussion}
\label{Concl}

In this article we have developed exhaustive search algorithms to find perturbatively flat vacua. The $2$-moduli case has been discussed in detail and an algorithm applicable to any number of moduli has been presented. Detailed numerical scans going through specific models (including ones with higher
number of complex structure moduli) will be presented elsewhere \cite{future1}. 

In Sec. \ref{Sec3} we have also examined perturbatively flat vacua as part of the entire ensemble of vacua at low $|W_0|$ from the point of view of a statistical approach. We found that they are statistically sparse when compared to the expectation from the distribution of low values of $|W_0|$ from \cite{Denef:2004ze}. This expectation has been confirmed in Sec. \ref{Sec4} by our numerical searches for cases with $2$ complex structure moduli. In particular, for the $\mathbb{CP}_{[1,1,1,6,9]}$ model we found that there is only $1$ perturbatively flat solution with $|W_0| \lesssim 0.01$ (featuring $|W_0|\sim 10^{-8}$), while \cite{Denef:2004ze} would predict around $\mc{O}(10^8)$ flux vacua. We argued that similar considerations apply to all other $2$-moduli cases. We therefore conclude that this set by itself does not provide tuning freedom for phenomenological applications. Using the general algorithm outlined in Sec. \ref{Sec5}, it would be interesting in the future to perform a detailed search for cases with more than $2$ complex structure moduli \cite{future1}, although one expects them to be statistically sparse from the analysis of Sec. \ref{Sec3}. Let us just mention here that, as one goes to higher values of $h^{1,2}_-$, one can expect more solutions. However, the analysis of \cite{Denef:2004dm} implies that with higher values of $h^{1,2}_-$ the vacua in the large complex structure limit give a lower contribution to the statistics. This poses an interesting challenge for achieving statistical tuning in phenomenological applications. Furthermore, one can expect that the numerics required to obtain all vacua explicitly should become harder as one goes up in the number of complex structure moduli.

Let us also stress that our analysis in Sec. \ref{Sec3} relied heavily on the specific form of the vacuum expectation values of the complex structure moduli (equation \pref{Ulocus} which is specific to the vacua of \cite{Demirtas:2019sip}) but in principle there can be other families of vacua featuring $W=0$ at perturbative level. An interesting question is to develop diagnostic methods to study the statistical significance of the vacua of \cite{Demirtas:2019sip} in general. Let us touch upon this briefly. For all such vacua, the instanton effects that give $W_0$ a non-zero value would also be responsible for giving the perturbatively flat direction a mass. Thus a universal property is a modulus (in the subspace spanned by the complex structure moduli and the axio-dilaton) with a low mass, more specifically a mass proportional to a positive power of $|W_0|$. Given this, one can ask whether there is a correlation between low $|W_0|$ and a modulus of low mass. This question can be addressed by examining the bosonic mass matrix of Sec. 3.2 of \cite{Denef:2004ze} (for which we refer the reader to the paper, as when presented in full glory the formulae are quite involved and we will only need some general features of the matrix for our discussion). If one of the masses scales as $|W_0|^k$ (for some positive $k$), then the determinant of the mass matrix would scale as $|W_0|^{2k}$, i.e. it would vanish in the $W_0 \to 0$ limit. On the other hand, taking $W_0 \to 0$ (which is equal to $X$ in the notation of \cite{Denef:2004ze}) is not a sufficient condition for the vanishing of the determinant. This indicates that the correlation is not universal, and so that there should exist another set of vacua with $W=0$ at perturbative level but with no flat directions. This observation agrees with the analysis of \cite{Burgess:2020qsc} based on scale invariance of the $10$-dimensional tree-level type IIB action. One family of the two original scaling symmetries is broken spontaneously by the vacuum expectation value of the dilaton, resulting in a massless Goldstone boson in $4$ dimensions which can be identified with $\tau$. Non-zero background fluxes can act as explicit symmetry breaking parameters (like non-zero quark masses in chiral perturbation theory), and can lift this flat direction. However, $W=0$ is not enough to guarantee no explicit breaking, and so no flat direction, since also derivatives of $W$ should vanish (see \cite{future2} for a study of flat directions in toroidal flux vacua in this context). 

Let us close by pointing out that, given a model, the statistical significance of any family of perturbatively flat vacua can be determined by the cut in the integration range of the flux variable $Z^I$ (of \cite{Denef:2004ze}) put by the requirement of a low mass (at  $|W_0|$ below a certain value). We hope to return to this question in the future.

\vspace{0.5 cm}
 
\section*{Acknowledgments}

We would like to thank Manki Kim, Sven Krippendorf, Liam McAllister, Ashoke Sen and Ravindranathan Thangadurai for useful discussions. AM is supported in part by the SERB, DST, Government of India by the grant MTR/2019/000267. The work of K. Sinha is supported in part by DOE Grant desc0009956.

\vspace{0.5cm}

\appendix

\section{$2$-moduli examples data}
\label{AppA}
 
As discussed in Sec. \ref{Sec5}, our numerical analysis has shown that in $22$ of the $39$ $2$-moduli examples in the Kreuzer-Skarke list, $|\epsilon|_{\rm inf}$ does not take values close to zero. This by itself gives a strong bound on the region where all possible solutions to the Lemma are contained (without the need of imposing requirements such as validity of the effective field theory or phenomenological viability). We list these models in Tab. \ref{Tab3} together with the associated values of $|\epsilon|_{\rm inf}$ and the values of $\vec{m}$ and $\vec{k}$ at which the infimum is attained.

\begin{table}
\centering
\begin{tabular}{ | p{2.0cm} | p{3.5cm}  | p{3.5cm} |p{2.5cm} | }
\hline
Model &   $ \vec{m}^T$ & $\vec{k}^T$& $|\epsilon|_{\rm inf}$\\
\hline
$M_{2,2}$ & $(-0.186, 0.982)$& $(-0.399,0.917)$ & $0.9751$\\
 \hline
$M_{2,3}$ & $(-0.966,0.257)$& $(-0.746,0.666)$ & $0.8919$\\ 
\hline
$M_{2,4}$ & $(-0.966,0.257)$& $(-0.746,0.666)$ & $0.8919$\\  
\hline
$M_{2,5}$ &  $(-0.966,0.257)$& $(-0.746,0.666)$ & $0.8919$\\ 
\hline
$M_{2,7}$ &  $(-0.967., 0.253)$& $(-0.778,0.628)$ & $0.9114$ \\
\hline
$M_{2,13}$ & $(-0.143, 0.989)$& $(-0.297, 0.955)$ & $0.9870$\\
 \hline
$M_{2,18}$ & $(-0.132, 0.991)$& $(-0.272, 0.962)$ & $0.9897$\\
\hline
$M_{2,19}$ & $(-0.186, 0.982)$& $(-0.399, 0.917)$ & $0.9750$\\
\hline
$M_{2,21}$ &$(0.966, -0.257)$& $(0.746,-0.666)$ & $0.8919$  \\
\hline
$M_{2,22}$ &$(-0.896, 0.438)$& $(-0.695,0.709)$ & $0.9336$  \\
\hline
$M_{2,23}$ &$(-0.896, 0.438)$& $(-0.695,0.709)$ & $0.9336$  \\
\hline
$M_{2,24}$ &$(-0.896, 0.438)$& $(-0.695,0.709)$ & $0.9336$  \\
\hline
$M_{2,25}$ &$(-0.186, 0.982)$& $(-0.399,0.917)$ & $0.9752$  \\
\hline
$M_{2,26}$ &$(-0.969, 0.243)$& $(-0.816,0.577)$ & $0.9321$  \\
\hline
$M_{2,27}$ &$(-0.969, 0.247)$& $(-0.599,0.801)$ & $0.7784$  \\
\hline
$M_{2,28}$ &$(-0.969, 0.247)$& $(-0.599,0.801)$ & $0.7784$  \\
\hline
$M_{2,29}$ &$(-0.969, 0.247)$& $(-0.599,0.801)$ & $0.7784$  \\
\hline
$M_{2,35}$ &$(-0.993, 0.114)$& $(-0.972,0.233)$ & $0.9927$  \\
\hline
$M_{2,36}$ &$(-0.186, 0.982)$& $(-0.399,0.917)$ & $0.9752$  \\
\hline
$M_{2,37}$ &$(-0.186, 0.982)$& $(-0.399,0.917)$ & $0.9752$  \\
\hline
$M_{2,38}$ &$(-0.186, 0.982)$& $(-0.399,0.917)$ & $0.9752$  \\
\hline
$M_{2,39}$ & $(-0.970, 0.243)$& $(-0.577, 0.817)$ & $0.7581$\\ 
\hline
\end{tabular}
\caption{$22$ $2$-moduli examples from the Kreuzer-Skarke list in which $|\epsilon|_{\rm inf}$ does not take values close to zero. For each model, we show the associated values of $|\epsilon|_{\rm inf}$ and the values of $\vec{m}$ and $\vec{k}$ at which the infimum is attained.}
\label{Tab3}
\end{table}

\end{document}